\documentclass[12pt]{article}
\usepackage{graphics,epsfig,wasysym,subfigure}

\begin{document}

\centerline{\large\bf Tension, rigidity and preferential curvature of}
\centerline{\large\bf interfaces between coexisting polymer solutions}
\vskip 20pt
\centerline{R.H. Tromp$^{1,2}$ and E.M. Blokhuis$^3$}
\vskip 10pt
\centerline{\it $^1$NIZO food research, Kernhemseweg 2, 6718 ZB Ede, The Netherlands}
\vskip 5pt
\centerline{\it $^2$University of Utrecht, Department of Chemistry,}
\centerline{\it Padualaan 8, 3584 CH Utrecht, The Netherlands}
\vskip 5pt
\centerline{\it $^3$Colloid and Interface Science, Leiden Institute of Chemistry,}
\centerline{\it Gorlaeus Laboratories, P.O. Box 9502, 2300 RA Leiden, The Netherlands}
\vskip 15pt
\centerline{\bf Abstract}
\vskip 5pt
\noindent
The properties of the interface in a phase-separated solution of polymers with different
degrees of polymerization and Kuhn segment lengths are calculated. The starting point is
the planar interface, the profile of which is calculated in the so-called `blob model', which
incorporates the solvent in an implicit way. The next step is the study of a metastable droplet
phase formed by imposing a chemical potential different from that at coexistence.
The pressure difference across the curved interface, which corresponds to this higher chemical
potential, is used to calculate the curvature properties of the droplet. Interfacial
tensions, Tolman lengths and rigidities are calculated and used for predictions for a realistic
experimental case. The results suggest that interfaces between phase-separated solutions of
polymers exhibit, in general, a preferential curvature, which stabilizes droplets of low molecular
mass polymer in a high molecular mass macroscopic phase.

\vskip 5pt
\noindent
\centerline{\rule{300pt}{1pt}}

\section{Introduction}
\label{sec-introduction}

\noindent
When two chemically different liquid polymers are mixed, in the majority of cases phase
separation takes place, because of the low gain in entropy after mixing, easily overruled by
an adverse enthalpy of mixing. The presence of a common solvent usually suppresses the
drive to phase separate, but only when the solvent is present at volume fractions of
typically 90\% or more. When the solvent is present at a lower concentration, phase
separation takes place, giving rise to phase regions differing in polymer composition
and total polymer concentration. The driving force for the phase separation is a repulsive,
possibly entropic, force between the polymers, or a difference in the quality of the solvent
for the two polymers \cite{Bergfeldt}. Phase separation in a polymer mixture can also occur
such that the polymers reside in one phase and the solvent in the other (complex coacervation).

Interfaces between phases of coexisting, thermodynamically incompatible polymer solutions
are referred to as {\em solvent-solvent} interfaces. An example is phase-separated
protein and (neutral) polysaccharide solutions found in food systems \cite{Tolstoguzov}.
The solvent-solvent interface, in that case a water-water interface, is then situated
between the coexistent protein-rich and polysaccharide-rich solutions.
The interfacial tension of solvent-solvent interfaces is extremely low (typically
a few $\mu$N/m or less) so that the interface is highly deformable, e.g. by convective flows,
and therefore difficult to investigate by classical methods (such as the Wilhelmy plate method).
Methods to estimate the interfacial tension are capillary rise \cite{Aarts},
spinning drop \cite{Ryden, deHoog, Scholten02, Scholten04}, shape relaxation after
deformation \cite{Simeone, vanPuyvelde, Antonov}, or by manipulation with a laser
beam \cite{Mitani}.

\begin{figure}
\centering
\includegraphics[width=230pt]{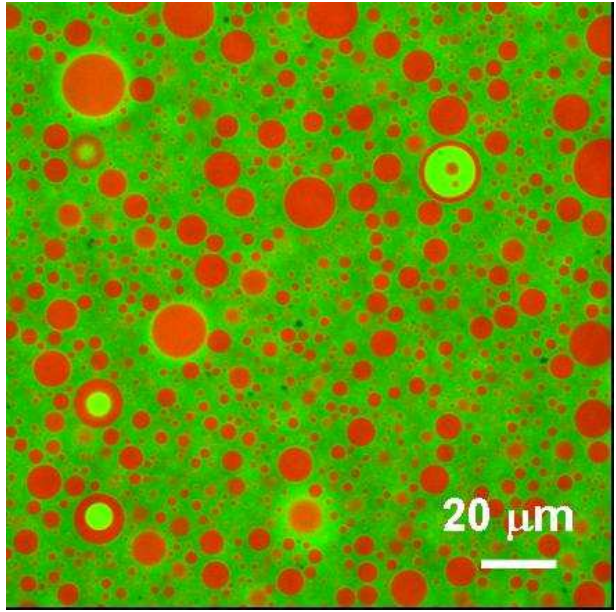}
\caption{Water-water emulsion, from a phase separated aqueous solution of 10\% fish gelatin
(non-gelling), and 10\% dextran (200 kDa). Light phase: gelatin-rich; dark phase: dextran-rich.}
\label{Fig:fig1}
\end{figure}

The macroscopic phase separation takes place via a stage in which one phase is a
dispersion of microscopic droplets in a continuous matrix of the other phase. This
state might be called a {\em solvent-solvent emulsion}. Figure~\ref{Fig:fig1} gives
an example. Such an emulsion is often surprisingly stable. This may be because of
the small density difference (both phases are typically 90\% solvent), low interfacial
tension or the existence of a preferential curvature. In order to get a basic theoretical
understanding of the properties that play a role in the stability of such solvent-solvent
emulsions, we present calculations of the interfacial tension, preferential curvature
and rigidity of equilibrium and quasi-equilibrated curved interfaces between polymer
phases. This is done using the so called 'blob model' \cite{DeGennes}. The blob model
is based on a Flory-Huggins mean-field approach but it is adapted to include excluded
volume fluctuations which are correlated over a blob size (correlation length) $\xi$.
Since solvent-solvent interfaces are wide, it contains only weak gradients so that it
seems appropriate to describe the interfacial region in terms of the van der Waals
squared-gradient expansion.

In this article, we discuss two simplified cases: (i) identical degrees of polymerization
and solvent qualities, but allowing for a solvent gradient in the interface,
(ii) different degrees of polymerization and solvent qualities (i.e. different Kuhn lengths),
but neglecting the solvent gradient in the interface. Case (i) was previously treated by
Broseta {\em et al.} \cite{Broseta86, Broseta87a, Broseta87b} and will be used as an
introduction to case (ii). A preferential curvature is expected in case (ii) because
the energy in the compositional gradient depends on the blob size ratio and the ratio
of the number of blobs per chain. 

\section{The blob model}

\subsection{Approach by Broseta {\em et al.}}

\noindent
The derivation of Broseta {\em et al.} \cite{Broseta86, Broseta87a, Broseta87b, Tromp}
starts with the Flory Huggins model \cite{FH} for an isotropic polymer
melt consisting of two polymer types of equal length $N$.
We denote the volume fractions of type A and type B as
$\phi_A \!=\! \phi$ and $\phi_B \!=\! 1- \phi$. Then,
\begin{equation}
\frac{F_{\rm FH}(\phi)}{V \, k_{\rm B} T} =  \frac{1}{a^3}
\left[ \frac{\phi}{N} \, \ln(\phi) + \frac{(1 - \phi)}{N} \, \ln(1-\phi)
+ \chi \, \phi \, (1 - \phi) \right] \,,
\end{equation}
where $a$ is the dimension of the lattice which equals the Kuhn segment
length. The Flory parameter $\chi$ represents the interaction between
the two polymer types.

The Flory-Huggins model is used by Broseta {\em et al.} \cite{Broseta87a}
for a mixture of two polymers well above the overlap concentration
(semi-dilute solution) in a solvent which is a good solvent for both polymers.
In that situation the {\bf blob model} can be used to describe the polymer
solution \cite{DeGennes}. The blob model incorporates the effect of
fluctuations due to excluded volume interactions by replacing the
Kuhn segment length $a$ by a `blob size' $\xi$ and the polymer length
$N$ by the number of blobs, $N_b \!=\! N / (c \, \xi^3)$ per polymer chain:
\begin{equation}
\frac{F_{\rm blob}(\phi,c)}{V \, k_{\rm B} T} =  \frac{1}{\xi^3}
\left[ \frac{\phi}{N_b} \, \ln(\phi) + \frac{(1 - \phi)}{N_b} \, \ln(1-\phi)
+ u \, \phi \, (1 - \phi) + K \right] \,,
\end{equation}
where $K \!=\! 0.024$ is a constant \cite{Broseta86, Broseta87b}, related
to the osmotic pressure, and where $u$ represents the interaction between
blobs of type A and B. The whole volume is now filled by `blobs' of dimension
$\xi$ with each `blob' either containing polymer of type A or type B.
An essential feature in the blob model is that both the blob size
$\xi \!=\! \xi(c)$ and the interaction parameter $u \!=\! u(c)$ depend
on the (total) {\em monomer concentration} $c$ \cite{DeGennes, Lapp}:
\begin{equation}
\label{eq:xi}
\xi(c) \simeq 0.43 \, R_{\rm g} \left( \frac{c}{c^*} \right)^{- \nu / (3 \nu -1)} \,,
\end{equation}
and
\begin{equation}
\label{eq:u}
u(c) = u_{\rm crit} \left( \frac{c}{c_{\rm crit}} \right)^{\chi / (3 \nu - 1)} \,,
\end{equation}
where the exponents $\nu \!=$ 0.60 and $\chi \!=$ 0.22 are such that
non-mean-field chain statistics is explicitly taken into account ($\chi$
can be interpreted as the number of monomeric contacts between overlapping blobs).
Furthermore, $c^*$ is the overlap concentration, $R_{\rm g}$ the polymer
radius of gyration in dilute (non-overlapping) conditions, and
$u_{\rm crit}$ is the interaction at a reference concentration,
for which we shall take the critical demixing concentration, $c_{\rm crit}$.

The blob model for an isotropic polymer mixture can be extended to
{\em non-homogeneous} polymer solutions. This is achieved by
allowing $\phi \!=\! \phi(\vec{r})$ and $c \!=\! c(\vec{r})$ to be
position dependent and by adding squared gradient terms to the free energy:
\begin{eqnarray}
\label{eq:F0}
\frac{F[\phi,c]}{k_{\rm B} T} &=& \int \!\! d\vec{r}
\left[ \frac{\phi}{N_b \, \xi^3} \, \ln(\phi) + \frac{(1 - \phi)}{N_b \, \xi^3} \, \ln(1-\phi)
+ \frac{u}{\xi^3} \, \phi \, (1 - \phi) \right. \nonumber \\
&& \left. + \frac{K}{\xi^3} + \frac{|\vec{\nabla} \phi|^2}{24 \, \xi \, \phi}
+ \frac{|\vec{\nabla} \phi|^2}{24 \, \xi \, (1-\phi)} + \frac{|\vec{\nabla}c|^2}{24 \, \xi \, c^2} \right] \,.
\end{eqnarray}
This expression is used by Broseta {\em et al.} \cite{Broseta87a}
to determine density profiles and surface tension of a {\em planar}
interface. We revisit their analysis in the Appendix and point
to some small errors in their original formulas.

We can extend the above free energy to {\em non-symmetric} polymer solutions
by allowing the blob size and number of blobs to differ for each polymer type.
This may be due to a different degree of polymerization, different solvent
qualities, different Kuhn lengths, etc. The difference in blob size ($\xi_A$
and $\xi_B$) and number of blobs ($N_{b,A}$ and $N_{b,B}$), then leads to the
following modification of the expression for the free energy:
\begin{eqnarray}
\label{eq:F1}
\frac{F[\phi,c]}{k_{\rm B} T} &=& \int \!\! d\vec{r}
\left[ \frac{\phi}{N_{b,A} \, \xi_A^3} \, \ln(\phi) + \frac{(1 - \phi)}{N_{b,B} \, \xi_B^3} \, \ln(1-\phi)
+ \frac{u}{\xi^3} \, \phi \, (1 - \phi) \right. \nonumber \\
&& \left. + \frac{K}{\xi^3} + \frac{|\vec{\nabla} \phi|^2}{24 \, \xi_A \, \phi}
+ \frac{|\vec{\nabla} \phi|^2}{24 \, \xi_B \, (1-\phi)} + \frac{|\vec{\nabla}c|^2}{24 \, \xi \, c^2} \right] \,,
\end{eqnarray}
where we have defined an effective blob size as
\begin{equation}
\label{eq:xi_eff}
\frac{1}{\xi} = \frac{1}{2} \left( \frac{1}{\xi_{A}} + \frac{1}{\xi_{B}} \right) \,.
\end{equation}
The form of the free energy in Eq.(\ref{eq:F1}) is derived by replacing
$\xi \!\rightarrow\! \xi_{A/B}$ and $N_b \!\rightarrow\! N_{b,A/B}$
in the corresponding terms in the free energy in Eq.(\ref{eq:F0}).
For the other terms in Eq.(\ref{eq:F0}), such as the term describing
the interaction between blobs, the blob size is replaced by an
{\em effective} blob size which should somehow be related to the two
blob sizes $\xi_A$ and $\xi_B$. We have chosen the expression in
Eq.(\ref{eq:xi_eff}) to calculate the effective blob size since it
reduces to $\xi$ for the symmetric polymer system and because its 
value is dominated by the {\em smallest} of the two blob sizes.
Naturally, other models for the free energy may be constructed
and, in particular, one may propose a different expression to
calculate the effective correlation length, but we expect that
most of the physics involved is captured by the expression
for the free energy in Eq.(\ref{eq:F1}). 

In the following sections, this free energy is applied to study spherically
(and cylindrically) shaped droplets. In the next section, we assume
that we can neglect the variation of the total monomer concentration
profile in the interfacial region and assume that $c(\vec{r}) \!=\! c$
everywhere.

\subsection{Calculation of the curved interface profile}

\noindent
When we neglect the variation of the total monomer concentration,
we have that $c(\vec{r}) \!=\! c$ and we may replace both the
blob size and the interaction parameter by their constant
values in the bulk region: $\xi \!\rightarrow\! \bar{\xi}$ and
$u \!\rightarrow\! \bar{u}$ (the bar denotes the corresponding bulk value).
Furthermore, the free energy is a functional of $\phi(\vec{r})$ only
and we can write for the grand free energy:
\begin{equation}
\label{eq:Omega}
\Omega[\phi] = \int \!\! d\vec{r} \left[ f(\phi) - \mu \, \phi + m(\phi) \, |\vec{\nabla} \phi|^2 \right] \,,
\end{equation}
where we have defined
\begin{equation}
f(\phi) = \frac{\bar{u} \, k_{\rm B} T}{\bar{\xi}^3}
\left[ \frac{2 \phi}{(1 + \alpha) \, \omega} \,
\ln(\phi) + \frac{2 \alpha \, (1 - \phi)}{(1 + \alpha) \, \omega} \, \ln(1 - \phi) - \phi^2 \right] \,,
\end{equation}
and
\begin{equation}
m(\phi) = \frac{k_{\rm B} T \, (1 + (r_0 - 1) \phi)}{12 \, \bar{\xi} \, (1 + r_0) \phi (1 - \phi)} \,,
\end{equation}
with the two asymmetry parameters $\alpha$ and $r_0$ defined as:
\begin{equation}
\label{eq:alpha}
\alpha \equiv \frac{\bar{N}_{b,A} \, \bar{\xi}_A^3}{\bar{N}_{b,B} \, \bar{\xi}_B^3} \,,
\end{equation}
and
\begin{equation}
\label{eq:r_0}
r_0 \equiv \frac{\bar{\xi}_A}{\bar{\xi}_B} \,.
\end{equation}
The parameter $\alpha$ represents the difference in degree of polymerization
of the two polymers, while $r_0$ reflects the difference in Kuhn lengths,
e.g. due to a difference in solvent quality or chain stiffness.
The dimensionless interaction parameter $\omega$ is defined by
\begin{equation}
\label{eq:omega_def}
\omega \equiv \frac{2 \bar{u}}{\bar{\xi}^3} \left( \frac{1}{\bar{N}_{b,A} \, \bar{\xi}_A^3}
+ \frac{1}{\bar{N}_{b,B} \, \bar{\xi}_B^3} \right)^{\!-1} \,,
\end{equation}
which can also be written as:
\begin{equation}
\label{eq:omega}
\omega = \frac{( 1 + \sqrt{\alpha} )^2}{1 + \alpha} \,
\left( \frac{c}{c_{\rm crit}} \right)^{(\chi + 1) / (3 \nu - 1)} \,,
\end{equation}
where we have used Eqs.(\ref{eq:u}) and (\ref{eq:alpha}) and the fact
that $\omega_{\rm crit} \!=\! ( 1 + \sqrt{\alpha} )^2 / (1 + \alpha)$.
This last expression will be used to link our theoretical results to
more experimentally accessible quantities.

Next, we consider a {\em spherically} shaped droplet with (equimolar) radius $R$.
The radius $R$ is really a {\em radius of curvature} and its sign is chosen
such that a {\em positive} curvature corresponds to the phase rich in polymer $A$
(the `liquid') residing {\em inside} the droplet with the (metastable) phase rich
in polymer $B$ (the `vapor') {\em outside}. As a consequence, $\alpha \!>\! 1$
means that the degree of polymerization inside the droplet is larger than outside,
and $r_0 \!>\! 1$ means that inside the droplet the characteristic length scale
(blob size) is larger than outside. The equimolar radius $R$ of a liquid drop
is defined in the expression:
\begin{equation}
\label{eq:R}
4 \pi \int\limits_{0}^{\infty} \!\! dr \; r^2
\left[ \, \phi(r) - \phi_v \right] = \frac{4 \pi}{3} \, R^3 \, (\phi_{\ell} - \phi_v) \,.
\end{equation}
The excess (grand) free energy of the metastable critical nucleus is defined as:
\begin{equation}
\frac{\Delta \Omega}{A} \equiv \frac{\Omega + p_v \, V}{A} =
- \frac{\Delta p \, R}{3} + \sigma(R) \,,
\end{equation}
with $p_v$ the vapour pressure outside the droplet and $p_{\ell} = p_v + \Delta p$
is the liquid pressure inside (see remark below, however). $A \!=\! 4 \pi \, R^2$
is the area of a droplet of radius $R$. The quantity $\sigma(R)$ is the surface
tension of the droplet as a function of the radius $R$. This is the quantity that we
wish to study and the above formula provides a way to determine it from $\Delta \Omega$.

\begin{figure}
\centering
\includegraphics[angle=270,width=230pt]{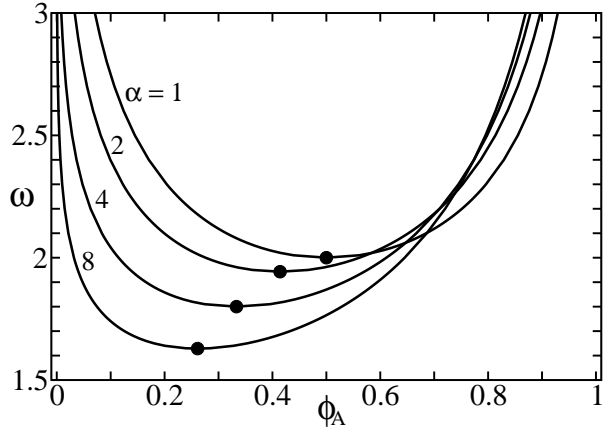}
\caption{Polymer phase diagram of mixing for several asymmetry ratios $\alpha$,
representing a difference in degree of polymerization (see Eq.(\ref{eq:alpha})).
$\omega$ is a dimensionless interaction parameter given in Eq.(\ref{eq:omega}).
The solid circles denote the locations of the critical demixing points,
$\omega_{\rm crit} \!=\! ( 1 + \sqrt{\alpha} )^2 / (1 + \alpha)$,
$\phi_{\rm crit} \!=\! 1 / ( 1 + \sqrt{\alpha} )$.}
\label{Fig:fig2}
\end{figure}

In spherical geometry, the free energy in Eq.(\ref{eq:Omega}) is:
\begin{eqnarray}
\label{eq:Delta_Omega}
\frac{\Delta \Omega[\phi]}{A} &=&  \int\limits_{0}^{\infty} \!\! dr \left( \frac{r}{R} \right)^{\!2}
\left[ f(\phi) - \mu \, \phi + p_v + m(\phi) \, \phi^{\prime}(r)^2 \, \right] \,.
\end{eqnarray}
The Euler-Lagrange equation to minimize the above free energy is given by
\begin{equation}
\label{eq:EL}
2 m(\phi) \, \phi^{\prime \prime}(r) = - \frac{4 m(\phi)}{r} \, \phi^{\prime}(r)
- m^{\prime}(\phi) \phi^{\prime}(r)^2 + f^{\prime}(\phi) - \mu \,,
\end{equation}
where the prime denotes a derivative with respect to its argument.
The procedure to determine $\sigma(R)$ as a function of $R$ is now as follows:
\vskip 5pt
\noindent
{\bf (1)} We first determine the bulk densities $\phi_{0,\ell}$ and $\phi_{0,v}$
coexisting across a planar interface ($R \!=\! \infty$), which we refer to as the
the state of coexistence. The corresponding value of the chemical potential is
then denoted as $\mu_{\rm coex}$. Determination of $\mu_{\rm coex}$, $\phi_{0,\ell}$
and $\phi_{0,v}$ is achieved by solving the following set of equations for
given $\omega$ and asymmetry parameters $\alpha$ and $r_0$:
\begin{eqnarray}
\label{eq:bulk_0}
f^{\prime}(\phi_{0,v}) &=& \mu_{\rm coex} \,, \hspace*{25pt}
f^{\prime}(\phi_{0,\ell}) = \mu_{\rm coex} \,, \nonumber \\
f(\phi_{0,v}) - \mu_{\rm coex} \, \phi_{0,v} &=& f(\phi_{0,\ell}) - \mu_{\rm coex} \, \phi_{0,\ell}
\equiv - p_{\rm coex} \,.
\end{eqnarray}
The resulting phase diagram is shown in Figure~\ref{Fig:fig2}
for various values of $\alpha$.
\vskip 5pt
\noindent
{\bf (2)} Next, we vary the chemical potential $\mu$ to a value {\em off-coexistence}.
The consequence of choosing a value of $\mu$ different from $\mu_{\rm coex}$ is the
coexistence of two pressures across a {\em curved} interface.
In the simplest case of spherical droplets we have droplets of liquid
($R \!>\! 0$) for $\mu \!>\! \mu_{\rm coex}$ and droplets of vapour ($R \!<\! 0$)
for $\mu \!<\! \mu_{\rm coex}$.
For given $\mu$, $\omega$ and $\alpha$ the densities $\phi_{\ell}$ and $\phi_v$
are determined from solving the following two equations:
\begin{equation}
f^{\prime}(\phi_{v}) = \mu \,, \hspace*{25pt} f^{\prime}(\phi_{\ell}) = \mu \,,
\end{equation}
with the corresponding bulk pressures calculated from
\begin{equation}
f(\phi_{v}) - \mu \, \phi_{v} = - p_{v} \,, \hspace*{25pt}
f(\phi_{\ell}) - \mu \, \phi_{\ell} = - p_{\ell} \,.
\end{equation}
It should be noted that far outside the droplet ($r \!\rightarrow\! \infty$)
the density (or pressure) becomes equal to that of the bulk,
$\phi(\infty) \!=\! \phi_{v}$, but that only for a very large droplet
is the density (or pressure) inside the droplet ($\phi(0)$)
equal to the bulk ($\phi_{\ell}$).
\vskip 5pt
\noindent
{\bf (3)} As a next step, the Euler-Lagrange equation for $\phi(r)$ in
Eq.(\ref{eq:EL}) is solved numerically with the boundary condition
$\phi(\infty) \!=\! \phi_{v}$. The resulting density profile
$\phi(r)$ thus obtained is finally inserted into Eq.(\ref{eq:R}) to
determine $R$ and into Eq.(\ref{eq:Delta_Omega}) to determine
$\Delta \Omega$ and thus $\sigma(R)$.

\begin{figure}
\centering
\includegraphics[angle=270,width=230pt]{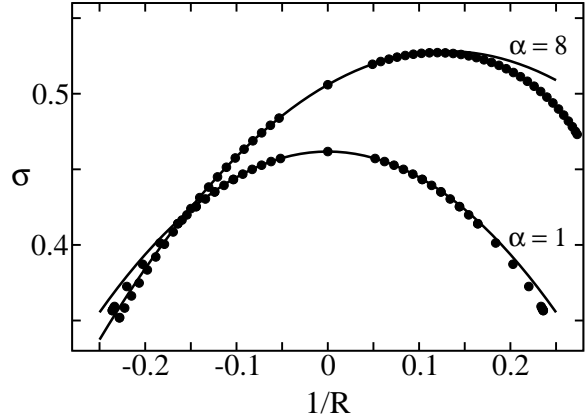}
\caption{Surface tension of a spherical droplet as function of
$1/R$, for ratios of degree of polymerization $\alpha =$ 1 and 8.
For the reduced units, see Eq.(\ref{eq:units}). $R \!>\! 0$
corresponds to droplets of the phase rich in polymer $A$
and $R \!<\! 0$ corresponds to droplets of the phase rich
in polymer $B$. We have set $\omega =$ 4 and $r_0 =$ 1 (equal
blob sizes). The solid lines are parabolic approximations to
$\sigma(R)$ determined from the curvature expansion.}
\label{Fig:fig3}
\end{figure}

In Figure~\ref{Fig:fig3}, two examples of $\sigma(R)$ are shown as a
function of $1/R$. These quantities are shown in reduced units which are
such that all energies are in units of $k_{\rm B} T / 6 (6 \bar{u})^{\frac{1}{2}}$
and all lengths are in units of $D_{\infty} \!\equiv\! \bar{\xi} / (6 \bar{u})^{\frac{1}{2}}$,
which is the interfacial width at infinite incompatibility of the two polymers,
i.e. at infinite degree of polymerization \cite{Broseta87a}.
This means that the surface tension and radius in reduced units are:
\begin{eqnarray}
\label{eq:units}
\tilde{\sigma}(R) &\equiv& \frac{\bar{\xi}^2}
{k_{\rm B} T \, (\bar{u}/6)^{\frac{1}{2}}} \, \sigma(R) \,, \\
\tilde{R} &\equiv& \frac{R}{D_{\infty}} = \frac{(6 \bar{u})^{\frac{1}{2}}}{\bar{\xi}} \, R \,. \nonumber
\end{eqnarray}
The determination of $\sigma(R)$ is quite elaborate and most of
the time we are only interested in its general shape. For this
reason we perform a curvature expansion in $1/R$ in the next
section to determine the parabolic approximation to $\sigma(R)$
shown as the solid line in Figure~\ref{Fig:fig3}.

\subsection{Curvature expansion}

\noindent
We can analyze the results in the previous section by performing a
curvature expansion in $1/R$. All quantities are then expanded in $1/R$.
For the density profile and surface tension, for instance, we have:
\begin{eqnarray}
\label{eq:expansion}
\phi(r)   &=& \phi_0(z) + \frac{\phi_1(z)}{R} + \ldots \,, \nonumber \\
\sigma(R) &=& \sigma - \frac{2 \delta \sigma}{R} + \frac{2 k + \bar{k}}{R^2} + \ldots \,,
\end{eqnarray}
where $z \!\equiv\! r - R$ and where $\sigma$ is the surface tension of the
planar interface, $\delta$ is the Tolman length \cite{Tolman}, related to
the radius of spontaneous curvature as $R_0 = 2 k / \delta \sigma$ \cite{Blokhuis92},
$k$ is the bending rigidity and $\bar{k}$ is the rigidity constant associated
with Gaussian curvature \cite{Helfrich}. An expansion of the Euler-Lagrange
equation in Eq.(\ref{eq:EL}) yields to zeroth order:
\begin{equation}
\label{eq:EL_0}
m(\phi_0) \, \phi_0^{\prime}(z)^2 = f(\phi_0) - \mu_{\rm coex} \, \phi_0(z) + p_{\rm coex} \,.
\end{equation}
Expansion of the Euler-Lagrange equation to first order gives:
\begin{eqnarray}
\label{eq:EL_1}
2 m(\phi_0) \, \phi_1^{\prime \prime}(z) &=& - 4 m(\phi_0) \, \phi_0^{\prime}(z) - \mu_1
- 2 m^{\prime}(\phi_0) \, \phi_0^{\prime}(z) \phi_1^{\prime}(z) \\
&& + \phi_1(z) \left[ f^{\prime \prime}(\phi_0)
- 2 m^{\prime}(\phi_0) \, \phi_0^{\prime \prime}(z)
- m^{\prime \prime}(\phi_0) \, \phi_0^{\prime}(z)^2 \, \right] \,. \nonumber
\end{eqnarray}
with $\mu_1 \!=\! 2 \sigma / (\phi_{0,\ell} - \phi_{0,v})$ \cite{Blokhuis93, Blokhuis06}.
One may show \cite{Blokhuis93} that the coefficients in the expansion of
$\sigma(R)$ in Eq.(\ref{eq:expansion}) can be written in terms of the
density profiles $\phi_0(z)$ and $\phi_1(z)$:
\begin{eqnarray}
\label{eq:curvature_coefficients_s}
\sigma &=& \int\limits_{-\infty}^{\infty} \!\!\! dz \; 2 m(\phi_0) \, \phi_0^{\prime}(z)^2 \,, \nonumber \\
\sigma \delta &=& - \int\limits_{-\infty}^{\infty} \!\!\! dz \; 2 m(\phi_0) \, z \, \phi_0^{\prime}(z)^2 \,, \\
2 k + \bar{k} &=& \int\limits_{-\infty}^{\infty} \!\!\! dz
\left[ 2 m(\phi_0) \, z^2 \, \phi_0^{\prime}(z)^2 - 2 m(\phi_0) \, \phi_1(z) \phi_0^{\prime}(z)
+ \mu_1 z^2 \, \phi_0^{\prime}(z) + \frac{\mu_1}{2} z \, \phi_1^{\prime}(z) \right] \,. \nonumber
\end{eqnarray}
The surface tension and Tolman length are independent of the choice
for the location of the interface (the position of the $z\!=\!0$ plane)
but $k$ and $\bar{k}$ {\em do} depend on this choice. For all of the
above expressions we have taken the {\em equimolar} radius for $R$,
see Eq.(\ref{eq:R}), which implies that the planar density profile obeys
\begin{equation}
\label{eq:equimolar}
\int\limits_{-\infty}^{\infty} \!\!\! dz \; z \, \phi_0^{\prime}(z) = 0 \,.
\end{equation}

The same expansion can be carried out for a {\em cylindrical} interface.
For the density profile and surface tension one then has instead of
Eq.(\ref{eq:expansion}):
\begin{eqnarray}
\label{eq:expansion_c}
\phi(r)   &=& \phi_0(z) + \frac{\phi_1(z)}{2 R} + \ldots \,, \nonumber \\
\sigma(R) &=& \sigma - \frac{\delta \sigma}{R} + \frac{k}{2 R^2} + \ldots \,.
\end{eqnarray}
This gives us the opportunity to determine $k$ and $\bar{k}$ separately
with the result \cite{Blokhuis93}
\begin{eqnarray}
\label{eq:curvature_coefficients_c}
\bar{k} &=& \int\limits_{-\infty}^{\infty} \!\!\! dz \;
2 m(\phi_0) \, z^2 \, \phi_0^{\prime}(z)^2 \,, \nonumber \\
k &=& \int\limits_{-\infty}^{\infty} \!\!\! dz
\left[ - m(\phi_0) \, \phi_1(z) \phi_0^{\prime}(z) + \frac{\mu_1}{2}
z^2 \, \phi_0^{\prime}(z) + \frac{\mu_1}{4} z \, \phi_1^{\prime}(z) \right] \,.
\end{eqnarray}
The procedure to determine the curvature coefficients $\sigma$,
$\delta$, $k$ and $\bar{k}$ is now as follows:
\vskip 5pt
\noindent
{\bf (1)} The planar profile $\phi_0(z)$ is first determined from
the differential equation in Eq.(\ref{eq:EL_0}) with
$\phi_{0,\ell}$, $\phi_{0,v}$, $\mu_{\rm coex}$ and $p_{\rm coex}$
determined from the set of equations in Eq.(\ref{eq:bulk_0}).
\vskip 5pt
\noindent
{\bf (2)} The location of the $z\!=\!0$ plane is chosen such that
Eq.(\ref{eq:equimolar}) is satisfied. This location
depends on the asymmetry of the interface profile.
\vskip 5pt
\noindent
{\bf (3)} With $\phi_0(z)$ determined, $\sigma$, $\delta$
and $\bar{k}$ can all be evaluated from the integrals in
Eqs.(\ref{eq:curvature_coefficients_s}) and (\ref{eq:curvature_coefficients_c}).
\vskip 5pt
\noindent
{\bf (4)} The constant $\mu_1$ is subsequently determined from
$\mu_1 \!=\! 2 \sigma / (\phi_{0,\ell} - \phi_{0,v})$ which
allows us to determine the bulk density values $\phi_{1,\ell/v}$
from $\phi_{1,\ell/v} \!=\! \mu_1 / f^{\prime \prime}(\phi_{0,\ell/v})$.
\vskip 5pt
\noindent
{\bf (5)} For given $\phi_0(z)$ and $\mu_1$, the differential equation
for $\phi_1(z)$ in Eq.(\ref{eq:EL_1}) may now be solved with the
boundary conditions $\phi_1(-\infty) \!=\! \phi_{\ell}$ and
$\phi_1(\infty) \!=\! \phi_{v}$. One may verify that if $\phi_1(z)$
is particular solution of Eq.(\ref{eq:EL_1}) then also
$\phi_1(z) + {\rm constant} \cdot \phi^{\prime}_0(z)$ is
also a solution. It turns out that $k$ is independent of the
value of this integration constant so we may arbitrarily 
take $\phi_1(0) \!=\! 0$ for convenience \cite{Blokhuis13}.
\vskip 5pt
\noindent
{\bf (6)} Finally, with $\phi_1(z)$ determined, $k$ (or $2 k + \bar{k}$)
can be evaluated from the integral in Eq.(\ref{eq:curvature_coefficients_c}).
\begin{figure}
\centering
\subfigure{\includegraphics[angle=270,width=200pt]{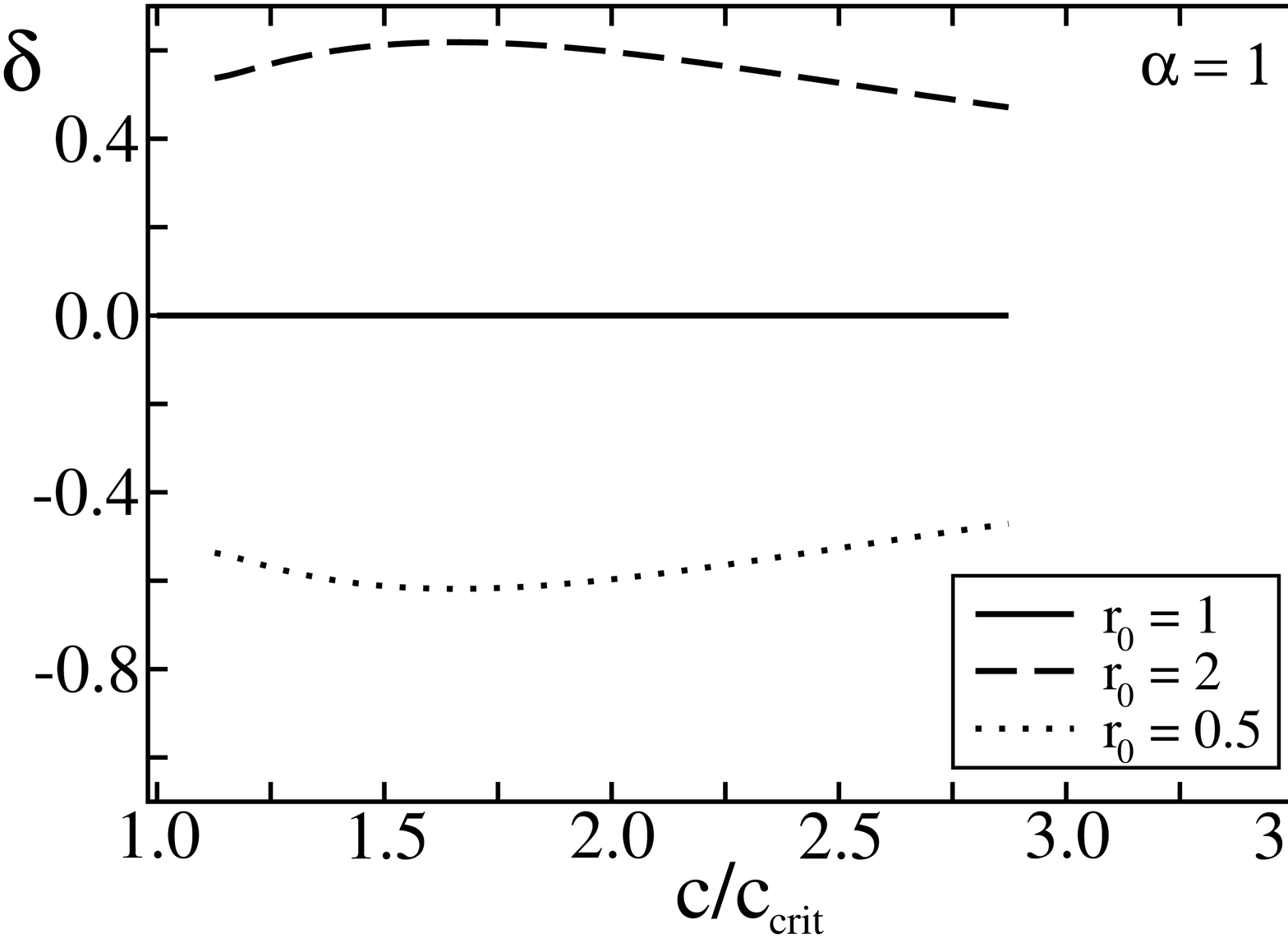}}
\subfigure{\includegraphics[angle=270,width=200pt]{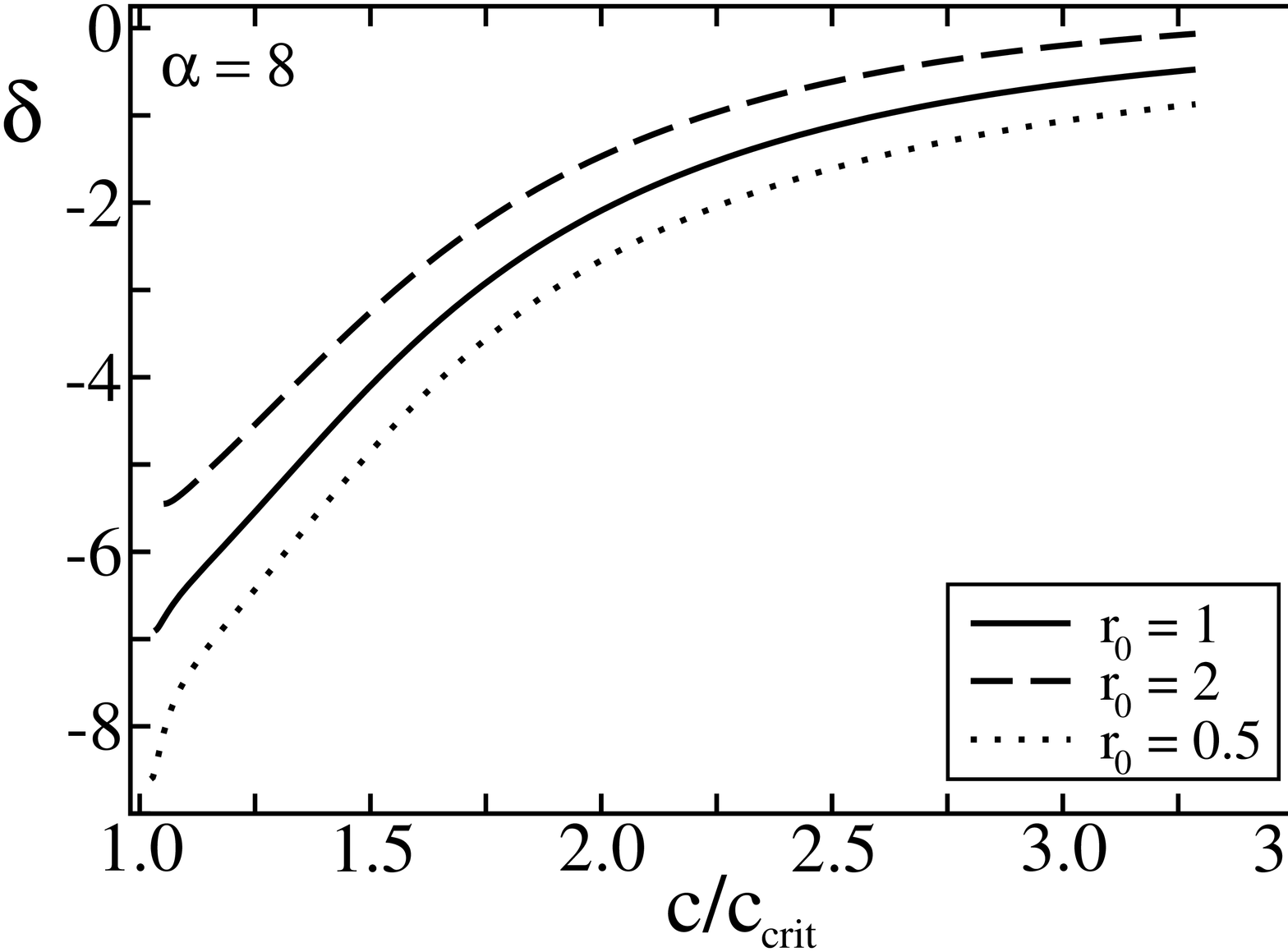}}\\
\caption{Variation of the Tolman length (in nm) as a function of
monomer concentration for two ratios of the degree of polymerization
and three ratios of the blob size.}
\label{Fig:fig4}
\end{figure}
\vskip 10pt
\noindent
The result of this procedure for the Tolman length is shown in
Figure~\ref{Fig:fig4} for various values of $\alpha$ and $r_0$.
For a perfectly symmetric profile ($\alpha \!=\! 1$ and $r_0 \!=\! 1$)
we have that $\delta \!=\! 0$ identically: there is no preferred curvature
to either phase. When $\alpha \!>\! 1$, we obtain {\em negative} values
for $\delta$ indicating that the droplets of the phase rich in A
($R \!>\! 0$) have a higher surface tension as compared to droplets
rich in polymer B ($R \!<\! 0$). This feature can also be seen from
the graph in Figure~\ref{Fig:fig3}. The influence of $r_0$ is less
pronounced with $r_0 \!>\! 1$ leading to a {\em positive} contribution
and $r_0 \!<\! 1$ leading to a {\em negative} contribution to $\delta$.

\begin{figure}
\centering
\subfigure{\includegraphics[angle=270,width=200pt]{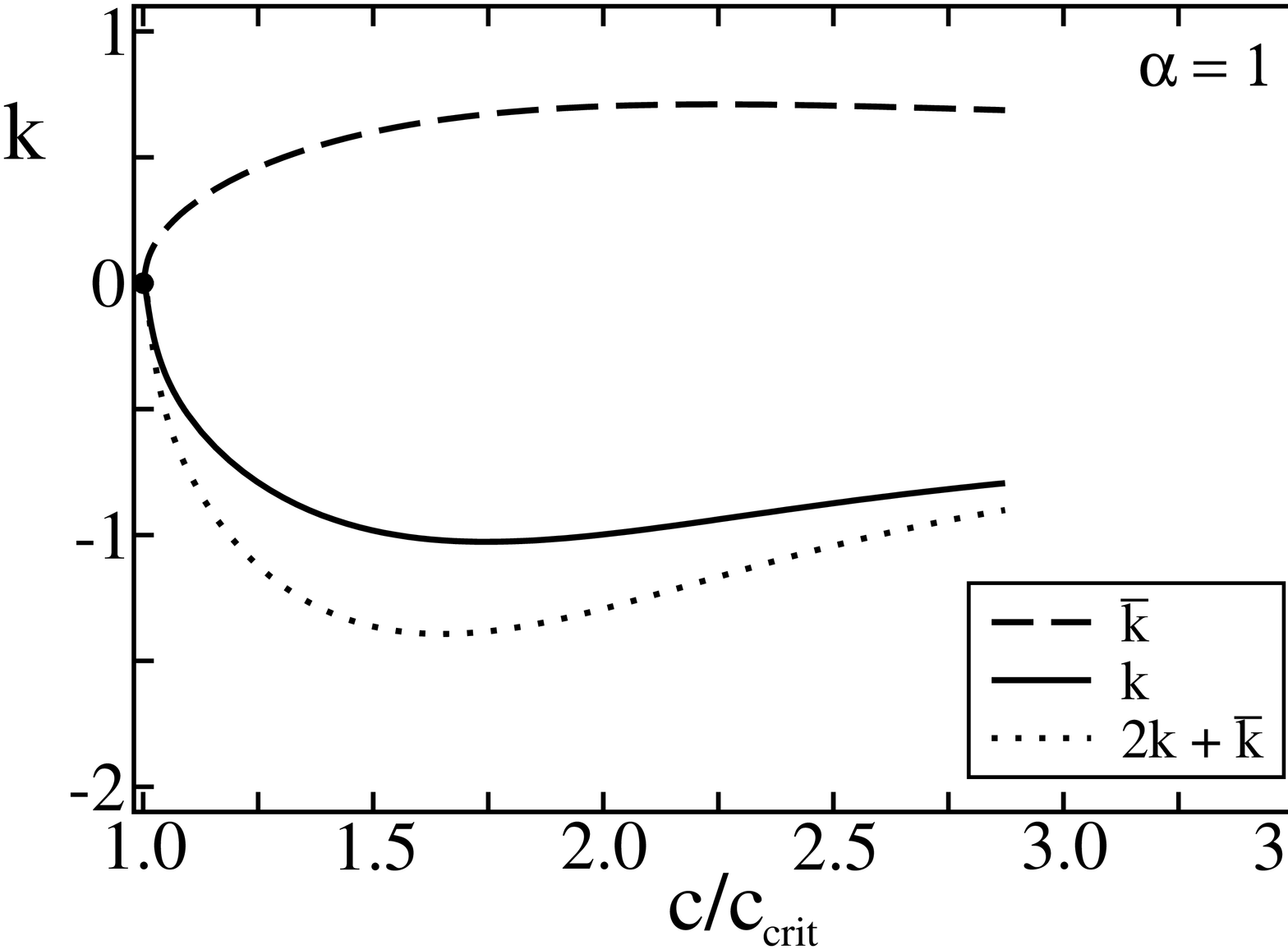}}
\subfigure{\includegraphics[angle=270,width=200pt]{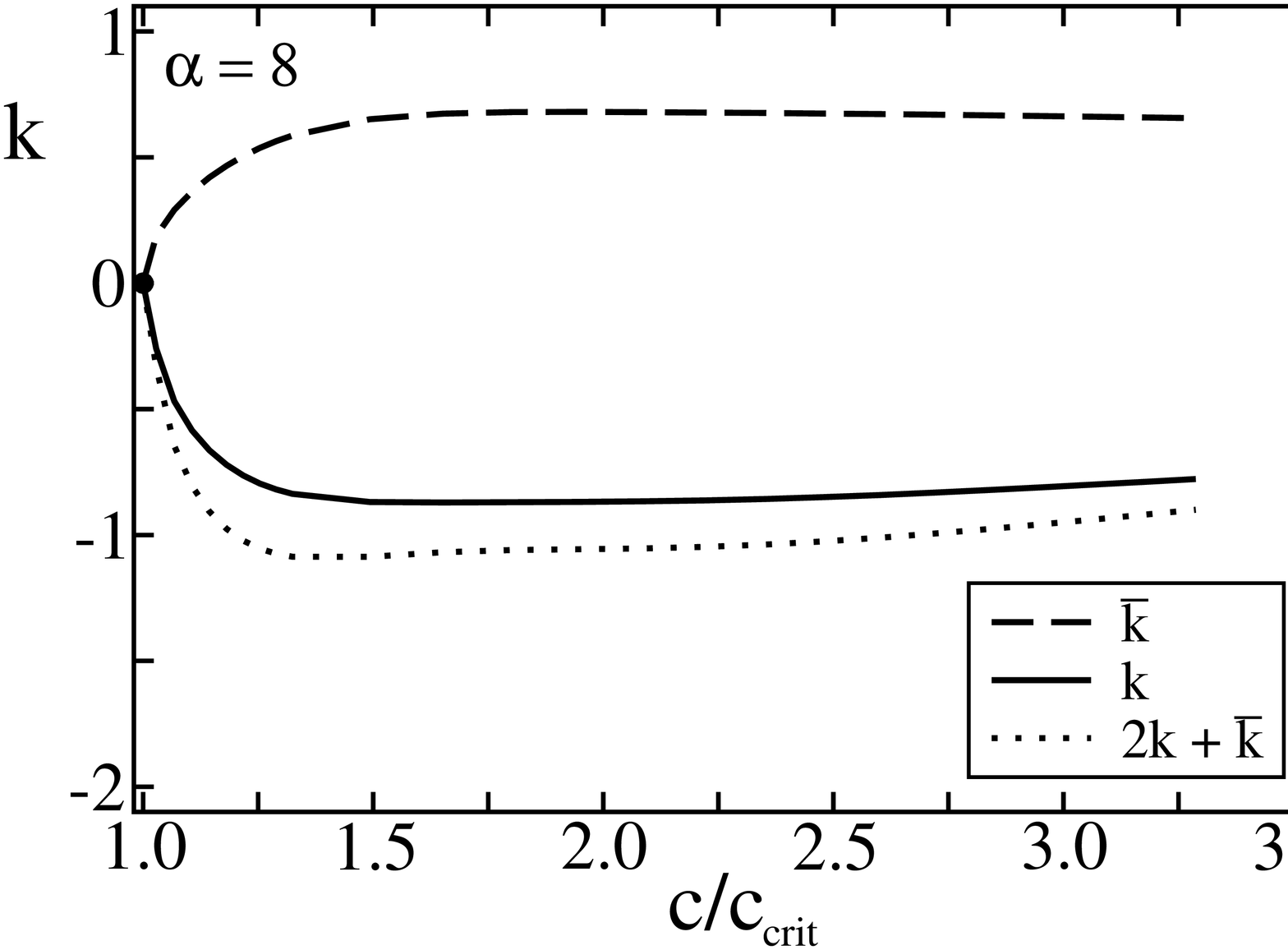}}\\
\caption{Variation of the rigidities (in units of $k_{\rm B} T$)
as a function of monomer concentration for two ratios of the
degree of polymerization. $k$ is the bending rigidity and
$\bar{k}$ is the rigidity constant associated with Gaussian
curvature.}
\label{Fig:fig5}
\end{figure}

In Figure~\ref{Fig:fig5}, we show the rigidity constants $k$ and $\bar{k}$
and the combination $2 k + \bar{k}$ in units of $k_{\rm B} T$ for
$\alpha \!=\! 1$ and for $\alpha \!=\! 8$. It is noted that
$2 k + \bar{k}$ is {\em negative} indicating that $\sigma(R)$ as a
function of $1/R$ has a {\em negative} second derivative, cf. Figure~\ref{Fig:fig3}.

\section{Discussion}

\noindent
In this section, we discuss in more detail the symmetric polymer system for
which the actual numbers quoted in Figures~\ref{Fig:fig4} and \ref{Fig:fig5}
are derived. The bar to denote bulk values shall be omitted in this section.
We first consider the symmetric, planar interface as treated by Broseta
{\em et al.} \cite{Broseta86, Broseta87a, Broseta87b}.

\subsection{Symmetric polymer system}

\noindent
For the symmetric polymer system, the influence of the solvent on the interfacial tension
of the planar interface can be investigated using the approach of Broseta {\em et al.}
\cite{Broseta86, Broseta87a, Broseta87b}, detailed in the Appendix. In order to
make the connection with experiments, we consider the parameters typical for
aqueous solutions of gelatin and dextran \cite{Tromp}, which have a radius of gyration
$R_{\rm g} \simeq$ 18 nm, degree of polymerisation $N \simeq$ 1000, and molecular
mass of the monomers, $M_{\rm mon} \simeq$ 120 g mol$^{-1}$. The overlap
concentration $c^*$ may then be estimated from
\begin{equation}
c^* \simeq \frac{N}{(4 \pi /3) R_{\rm g}^3} \,,
\end{equation}
which gives $c^* \simeq$ 0.82\% mass per mass solution. By convention, we shall
write concentrations in units of \% mass per mass solution which is achieved
by multiplying the concentration expressed as number of molecules per volume
by the molecular mass and dividing by Avogadro's number and the solvent (water)
mass density.

The blob size $\xi$ in this symmetric polymer system can be calculated as a
function of the bulk monomer concentration $c$ using Eq.(\ref{eq:xi}) \cite{Lapp}:
\begin{equation}
\label{eq:xi_2}
\xi(c) \simeq 0.43 \, R_{\rm g} \left( \frac{c}{c^*} \right)^{- \nu / (3 \nu - 1)} \,.
\end{equation}
From the blob size, the number of blobs per chain as a function of
the total monomer concentration is then determined from:
\begin{equation}
\label{eq:N_b}
N_{\rm b}(c) = \frac{N}{c \, \xi(c)^3} \,.
\end{equation}
For the polymer solution consisting of gelatin and dextran, the critical
demixing concentration is estimated as $c_{\rm crit} \simeq$ 3.5\% mass
per mass solution. The blob size at the critical concentration can then
be estimated from Eq.(\ref{eq:xi_2}), giving $\xi_{\rm crit} \simeq$ 2.6 nm,
and the number of blobs per chain at the critical demixing concentration
is estimated from Eq.(\ref{eq:N_b}) as $N_{\rm b, crit} \simeq$ 325.

By choosing in Eq.(\ref{eq:u}) the critical concentration of demixing
as the reference concentration, $u$ also becomes experimentally
accessible:
\begin{equation}
\label{eq:u_2}
u(c) = \frac{\omega_{\rm crit}}{N_{\rm b, crit}} \left( \frac{c}{c_{\rm crit}}
\right)^{\!\chi/(3 \nu - 1)} \,,
\end{equation}
where we have used that $\omega \!=\! N_{\rm b} u$ \cite{Broseta87a}.
For the symmetrical polymer system ($\alpha \!=\! 1$), we have that
$\omega_{\rm crit} \!=$ 2, which gives $u_{\rm crit} \!\simeq$ 0.0061.
In Table 1, we list the values of the parameters of the gelatin-dextran
system together with the parameters calculated from these numbers
and their range when the monomer concentration is varied between
$c \!=$ 3.5\% (w/w) and $c \!=$ 10\% (w/w).

\begin{table}
\centering
\begin{tabular}{|c|l|c|}
\hline
\multicolumn{2}{|c|}{quantity} & value \\
\hline
\hline
$R_{\rm g}$       & \hspace*{3pt} radius of gyration                 & 18 nm \\
$N$               & \hspace*{3pt} degree of polymerisation           & 1000 \\
$M_{\rm mon}$     & \hspace*{3pt} monomer molecular mass             & 120 g mol$^{-1}$ \\
\hline
$c^*$             & \hspace*{3pt} monomer overlap concentration      & 0.82\% (w/w) \\
\hline
\hline
$c_{\rm crit}$    & \hspace*{3pt} critical demixing concentration    & 3.5\% (w/w) \\
\hline
$\xi_{\rm crit}$  & \hspace*{3pt} critical blob size                 & 2.6 nm \\
\hspace*{3pt} $N_{\rm b, crit}$ \hspace*{3pt}
& \hspace*{3pt} critical number of blobs per chain \hspace*{3pt}     & 325 \\
$u_{\rm crit}$    & \hspace*{3pt} critical interaction parameter     & 0.0061 \\
\hline
\hline
$c$         & \hspace*{3pt} monomer concentration    & \hspace*{3pt} 3.5-10\% (w/w) \hspace*{3pt} \\
\hline
$\xi$       & \hspace*{3pt} blob size                 & 2.6-1.2 nm \\
$N_{\rm b}$ & \hspace*{3pt} number of blobs per chain & 325-1200 \\
$u$         & \hspace*{3pt} interaction parameter     & 0.0061-0.0084 \\
\hline
\end{tabular}
\caption{Values of the parameters of the gelatin-dextran system \cite{Tromp}
and their variation with monomer concentration $c$.}
\end{table}

\begin{figure}
\centering
\includegraphics[angle=270,width=200pt]{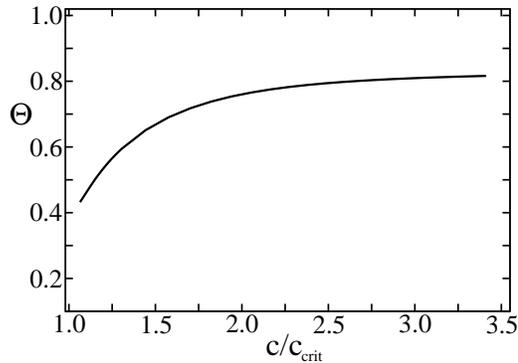}
\caption{Ratio of the interfacial tension corrected and not corrected
for the effect of solvent redistribution as a function of the (bulk)
monomer concentration. In this example, we have set $u \!=$ 0.05.}
\label{Fig:fig6}
\end{figure}

The free energy by Broseta {\em et al.} in Eq.(\ref{eq:F0}) for the symmetric polymer system
takes into account the reduction of the monomer concentration by the
accumulation of solvent at the interface, which is ignored in the calculation
of the properties of non-symmetric interfaces. It is therefore useful to
first consider the relevance of this effect. This can be done by plotting
the ratio of the interfacial tension of the planar interface calculated
with and without taking the reduction due to the solvent redistribution
into account. This ratio is (see Eq.(\ref{eq:A12})):
\begin{equation}
\Theta \equiv \frac{\sigma_{c = c(z)}}{\sigma_{c = \rm constant}} =
\frac{1 - \Delta_1 - u \, \Delta_2}{1 - \Delta_1} \,.
\end{equation}
It turns out that the effect of solvent redistribution is significant
in the whole concentration range, reducing the interfacial tension by
at least 20\% (See Figure~\ref{Fig:fig6}).
However, the reduction is nearly constant above a monomer concentration
of 1.5 $c_{\rm crit}$ ($\omega \apprge$ 3.7), above which the decreasing
amount of solvent is compensated by the increasing segregation of the
polymers at the interface. It is tentatively assumed that calculations
which ignore the solvent become more reliable at increasing monomer
concentration. The properties of the curved interface should therefore
be considered only at $\omega \apprge$ 4.

\subsection{Asymmetric polymer system}

\noindent
In our theoretical analysis, the Tolman length and rigidity constants are
calculated, as a function of the dimensionless interaction parameter
$\omega$, in reduced units similar to those in Eq.(\ref{eq:units}):
\begin{equation}
\tilde{\delta} \equiv \frac{\delta}{D_{\infty}} = \frac{(6 u)^{\frac{1}{2}}}{\xi} \, \delta
\hspace*{25pt} {\rm and} \hspace*{25pt}
\tilde{k} \equiv \frac{6 (6 u)^{\frac{1}{2}}}{k_{\rm B} T} \, k \,. \nonumber
\end{equation}
To transform these units into experimentally relevant values, we need to make
assumptions on the value of the effective blob size $\xi$ (Eq.(\ref{eq:xi_eff}))
and the blob interaction parameter $u$ (Eq.(\ref{eq:omega_def})) also for the
{\em asymmetric} polymer system. [The value and concentration dependence of $\omega$
is given by Eq.(\ref{eq:omega}).] Since the concentration dependence
of $\xi$ and $u$ is the same as in Eqs.(\ref{eq:xi_2}) and (\ref{eq:u_2}),
we therefore only need to assume values for their {\em critical} values 
$\xi_{\rm crit}$ and $u_{\rm crit}$. As a first guess, it seems
appropriate to assign for $\xi_{\rm crit}$ and $u_{\rm crit}$, the
same values as in the symmetric case, $\xi_{\rm crit} \!\simeq$ 2.6 nm
and $u_{\rm crit} \!\simeq$ 0.0061.

The Tolman length (Figure~\ref{Fig:fig4}) is, as obtained before \cite{Blokhuis06},
in the range of {\em nanometers}. The Tolman length is related to a preferential
curvature of the interface and it can differ from zero because of unequal degrees
of polymerization or because of unequal blob sizes.
For a prediction of the sign of the Tolman length for a specific pair of polymers,
one therefore needs to know the molecular mass ratio and the respective Kuhn lengths.

The implications of the sign of the Tolman length are apparent in Figure~\ref{Fig:fig3}
which shows the interfacial tension as a function of radius of curvature for a symmetric
and asymmetric polymer system. According to the definition of the asymmetry parameter
$\alpha$ in Eq.(\ref{eq:alpha}), the droplet phase has a higher degree of polymerization
than the surrounding infinite phase when $\alpha \!>\! 1$. Therefore, the shape of
$\sigma(R)$ as a function of $1/R$ in Figure~\ref{Fig:fig3} implies that in a
phase-separated polymer mixture, a droplet of {\em high} molecular mass has a
higher surface energy than a droplet of {\em low} molecular mass. This would
imply that the formation of droplets of the lower molecular mass is energetically
more favourable.

\subsection{Droplet stability}

\noindent
To elucidate the role of the radius dependent surface tension on the stability
of the droplets, we consider the free energy of a dispersion of $N_s$ spherical
droplets with radius $R$:
\begin{equation}
\Omega(N_s,R) = N_s \, 4 \pi R^2 \, \left[ \, \sigma - \frac{2 \delta \sigma}{R}
+ \frac{2 k + \bar{k}}{R^2} \, \right] -\Delta p \, N_s \, \frac{4 \pi}{3} R^3 \,.
\end{equation}
The above free energy should be minimized with respect $N_s$ and $R$ keeping
the total volume of the particles, $V_0 \!=\! \, N_s \, (4 \pi / 3) R^3$,
fixed. This yields the Laplace equation for the pressure difference:
\begin{equation}
\Delta p = \frac{2 \sigma}{R} - \frac{2 \delta \sigma}{R^2} \,,
\end{equation}
and we obtain for the {\em preferential radius}:
\begin{equation}
\label{eq:R_pref}
R_{\rm pref} =  2 \delta + \left( 4 \delta^2 - \frac{3}{\sigma} (2 k + \bar{k}) \right)^{\!\frac{1}{2}} \,.
\end{equation}
\begin{figure}
\centering
\includegraphics[angle=270,width=200pt]{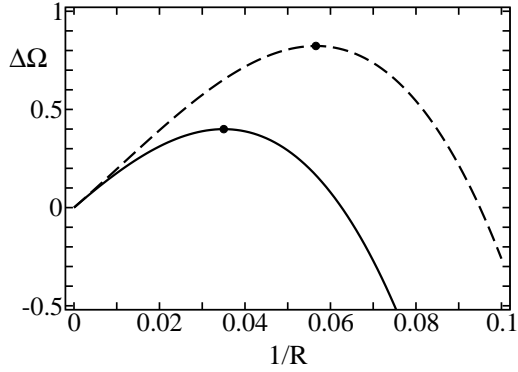}
\caption{Excess free energy density relative to the planar state $\Delta \Omega$
(in units of $k_{\rm B} T$/(10 nm)$^3$) as a function of the inverse radius
(in units of nm$^{-1}$). The solid line corresponds to droplets containing
low molecular mass polymer surrounded by a continuous phase consisting of
the high molecular mass polymers. The dashed line is the reversed situation.
The maximum (solid circle) in the free energy corresponds to the {\em preferential radius}
determined from Eq.(\ref{eq:R_pref}). In this example $\alpha =$ 8 and $c / c_{\rm crit} \!=$ 1.8.}
\label{Fig:fig7}
\end{figure}
The role of the preferential radius is illustrated in Figure~\ref{Fig:fig7}.
Using the parameters listed in Table 1, the excess free energy density relative
to the planar state, $\Delta \Omega \!\equiv\! \left[ \Omega(R) - \Omega(R=\infty) \right] / V_0$
is plotted as a function of the inverse radius $1/R$. The radius $R$ is the
droplet radius of the low molecular mass droplets (solid line) or the droplet
radius of the high molecular mass droplets (dashed line). In either case, the
maximum in the free energy corresponds to the {\em preferential radius} determined
from Eq.(\ref{eq:R_pref}). The shape of the free energy shows the existence of
metastable (and possibly long-lived) droplets with a preferential size that
ultimately grow to form a single phase ($R \!=\! \infty$). Figure~\ref{Fig:fig7}
also shows that the system's free energy is significantly lower for the formation
of droplets of low molecular mass polymer. This result indicates that when the
emulsion is initially formed, the thermoydnamic path followed toward macroscopic
phase separation is preferably through the formation of droplets containing polymer
with the {\em lower} molecular mass. This effect should be observable by dynamic
light scattering, when the two polymers have a high degree of monodispersity and
the phases have a small difference in density.

\begin{figure}
\centering
\subfigure{\includegraphics[angle=270,width=200pt]{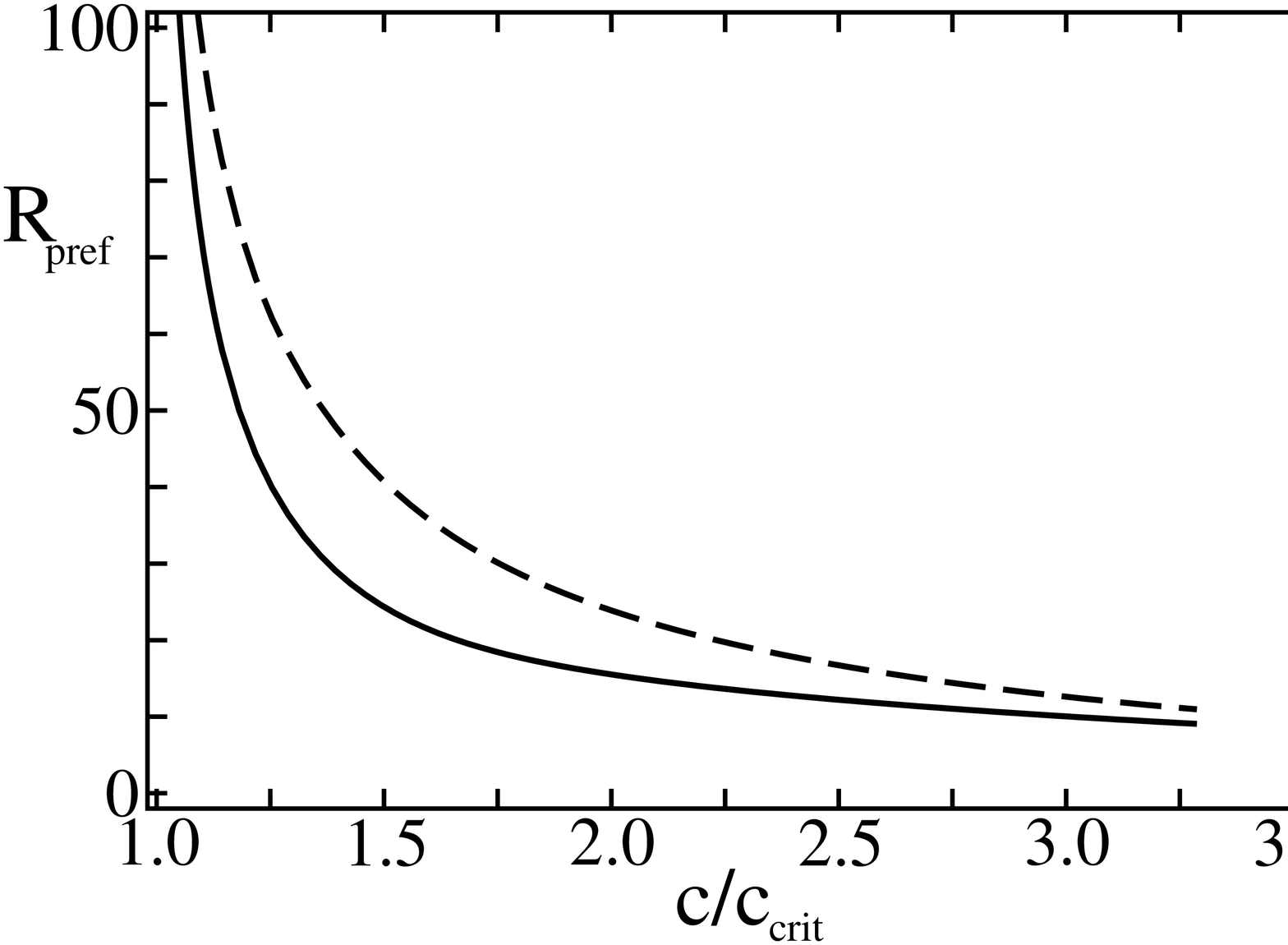}}
\subfigure{\includegraphics[angle=270,width=200pt]{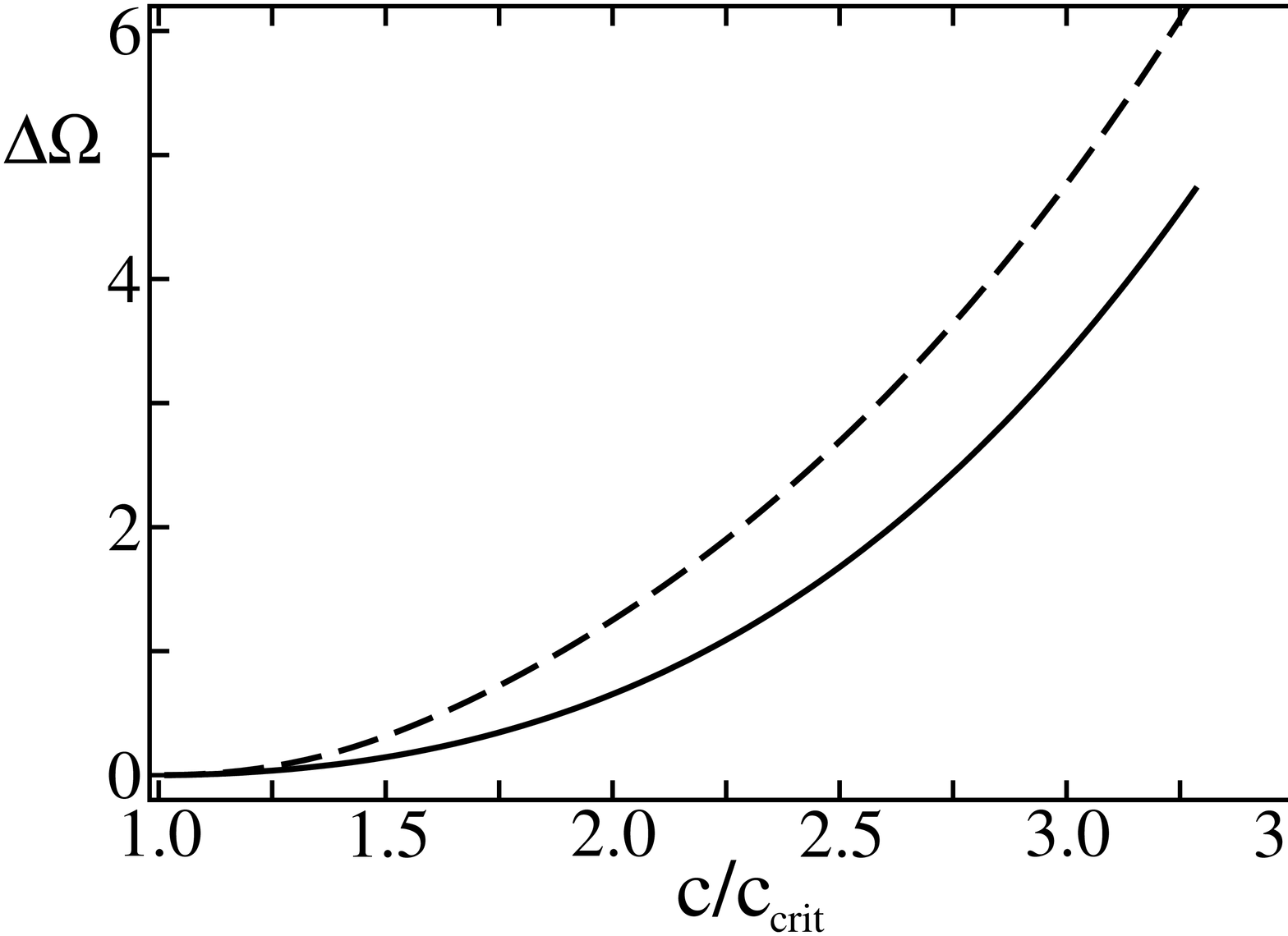}}\\
\caption{Preferential radius (in nm) and corresponding excess free energy density
relative to the planar state (in units of $k_{\rm B} T$/(10 nm)$^3$).
The solid line corresponds to droplets containing low molecular mass polymer
surrounded by a continuous phase consisting of the high molecular mass polymers.
The dashed line is the reversed situation. In this example $\alpha =$ 8.}
\label{Fig:fig8}
\end{figure}
The preferential radius and the corresponding free energy are shown as a function
of concentration in Figure~\ref{Fig:fig8}. The preferential size of the (metastable)
droplets is in the range of tenths of nanometers with the corresponding excess free
energy of the polymer system of the order of a few $k_{\rm B} T$ per volume of (10 nm)$^3$.

\section{Summary}

\noindent
The interfacial tension between coexisting incompatible polymer solutions is calculated
as a function of concentration for different degrees of polymerization and Kuhn segment
lengths. This is done using the blob model for semi-dilute polymer solutions. It turns
out that, when the degree of polymerization or the Kuhn length of the incompatible polymers
differs, the interfacial tension is lowest when the interface has a certain preferential
curvature. As a consequence, droplets of low molecular mass phase surrounded by a high
molecular mass bulk phase may have an energy which is lower than that of the reversed
situation. The preferential size was found to be of the order of tens of nanometers,
and therefore observable by light scattering.

\appendix

\section{Original analysis by Broseta}
\label{app}

\noindent
In this Appendix, we follow the analysis by Broseta {\em et al.}
\cite{Broseta86, Broseta87a, Broseta87b} of the profile of a planar
interface between two coexistent polymer phases with solvent.
The polymers have equal degrees of polymerization and are chemically
identical with respect to the solvent. We start with expression
for the free energy in Eq.(\ref{eq:F0}) for a planar interface:
\begin{eqnarray}
\frac{F[\phi,c]}{A \, k_{\rm B} T} &=& \int\limits_{-\infty}^{\infty} \!\!\! dz
\left[ \frac{\phi}{N_b \, \xi^3} \, \ln(\phi) + \frac{(1 - \phi)}{N_b \, \xi^3} \, \ln(1-\phi)
+ \frac{u}{\xi^3} \, \phi \, (1 - \phi) \right. \nonumber \\
&& \left. + \frac{K}{\xi^3} + \frac{\phi^{\prime}(z)^2}{24 \, \xi \, \phi}
+ \frac{\phi^{\prime}(z)^2}{24 \, \xi \, (1-\phi)} + \frac{c^{\prime}(z)^2}{24 \, \xi \, c^2} \right] \,.
\end{eqnarray}
Broseta {\em et al.} \cite{Broseta87a} continue by defining the dimensionless
$\eta(z)$ and $\varepsilon(z)$ as:
\begin{equation}
\phi(z) \equiv \frac{1 + \eta(z)}{2} \hspace*{25pt}
{\rm and} \hspace*{25pt} c(z) \equiv c + \bar{u} \, c \, \varepsilon(z) \,.
\end{equation}
The free energy per unit area then becomes
\begin{eqnarray}
\frac{F[\eta,\varepsilon]}{A \, k_{\rm B} T} &=& \int\limits_{-\infty}^{\infty} \!\!\! dz
\left[ \frac{(1+\eta)}{2 N_b \, \xi^3} \, \ln(1+\eta) + \frac{(1 - \eta)}{2 N_b \, \xi^3} \, \ln(1-\eta)
+ \frac{u}{4 \, \xi^3} \, (1 - \eta^2) \right. \nonumber \\
&& \left. + \frac{K}{\xi^3} + \frac{\eta^{\prime}(z)^2}{24 \, \xi \, (1 - \eta^2)}
+ \frac{\bar{u}^2}{24 \, \xi} \frac{\varepsilon^{\prime}(z)^2}{(1 + \bar{u} \varepsilon)^2} \right] \,.
\label{eq:A3}
\end{eqnarray}
Next, an expansion is made by Broseta {\em et al.} \cite{Broseta87a} assuming
that $\varepsilon(z)$ is small everywhere. One then has:
\begin{eqnarray}
\frac{1}{N_b \, \xi^3} &=& \frac{1}{\bar{N}_b \, \bar{\xi}^3} ( 1 + \bar{u} \varepsilon ) \,, \nonumber \\
\frac{1}{\xi^3} &=& \frac{1}{\bar{\xi}^3} ( 1 + \frac{3 \nu}{(3 \nu - 1)} \, \bar{u} \varepsilon
+ \frac{3 \nu}{2 \, (3\nu-1)^2} \, (\bar{u} \varepsilon)^2 + \ldots) \,, \nonumber \\
\frac{u}{\xi^3} &=& \frac{\bar{u}}{\bar{\xi}^3} ( 1 + \frac{(3 \nu + \chi)}{(3 \nu - 1)} \, \bar{u} \varepsilon + \ldots) \,.
\end{eqnarray}
To second order in $\varepsilon$, we then have for the free energy 
\begin{eqnarray}
\label{eq:F0_1}
\frac{F[\eta,\varepsilon]}{A \, k_{\rm B} T} &=& \frac{\bar{u}}{\bar{\xi}^3}
\int\limits_{-\infty}^{\infty} \!\!\! dz
\left[ \frac{(1+\eta)}{2 \omega} \, \ln(1+\eta) + \frac{(1 - \eta)}{2 \omega} \, \ln(1-\eta)
- \frac{\eta^2}{4} + \frac{\bar{\xi}^2 \, \eta^{\prime}(z)^2}{24 \, \bar{u} \, (1 - \eta^2)} \right. \nonumber \\
&& \left.+ \frac{(1+\eta)}{2 \omega} \, \ln(1+\eta) \, \bar{u} \, \varepsilon
+ \frac{(1 - \eta)}{2 \omega} \, \ln(1-\eta) \, \bar{u} \, \varepsilon 
- \frac{(3 \nu + \chi)}{4 \, (3 \nu - 1)} \, \eta^2 \, \bar{u} \, \varepsilon \right. \nonumber \\
&& \left. + \frac{3 \nu \, K}{2 \, (3 \nu - 1)^2} \, \bar{u} \, \varepsilon^2
+ \frac{\bar{\xi}^2}{24} \, \bar{u} \, \varepsilon^{\prime}(z)^2 \right] \,,
\end{eqnarray}
where we have defined $\omega \!=\! \bar{N}_b \, \bar{u}$ \cite{Broseta87a} and
omitted constants and terms proportional to a constant times $\varepsilon$.
These are adsorbed in a constant bulk pressure $\tilde{p}$ and chemical
potentials $\mu_{\eta}$ and $\mu_{\varepsilon}$. As a final step, we introduce
$x \!\equiv\! z / D_{\infty}$, with $D_{\infty} \!\equiv\! \bar{\xi} /
(6 \bar{u})^{\frac{1}{2}}$ and the excess grand free energy due to the interface
becomes:
\begin{eqnarray}
\label{eq:F0_2}
\frac{\Omega[\eta,\varepsilon]}{A \, k_{\rm B} T} &=& \frac{(\bar{u}/6)^{\frac{1}{2}}}{\bar{\xi}^2}
\int\limits_{-\infty}^{\infty} \!\!\! dx
\left[ \frac{(1+\eta)}{2 \omega} \, \ln(1+\eta) + \frac{(1 - \eta)}{2 \omega} \, \ln(1-\eta)
- \frac{\eta^2}{4} + \frac{\eta^{\prime}(x)^2}{4 \, (1 - \eta^2)} \right. \nonumber \\
&& \left. + \frac{(1+\eta)}{2 \omega} \, \ln(1+\eta) \, \bar{u} \, \varepsilon
+ \frac{(1 - \eta)}{2 \omega} \, \ln(1-\eta) \, \bar{u} \, \varepsilon 
- \frac{(3 \nu + \chi)}{4 \, (3 \nu - 1)} \, \eta^2 \, \bar{u} \, \varepsilon \right. \nonumber \\
&& \left. + \frac{3 \nu \, K}{2 \, (3\nu-1)^2} \, \bar{u} \, \varepsilon^2
+ \frac{\bar{u}^2}{4} \, \varepsilon^{\prime}(x)^2 - \mu_{\eta} \, \eta
- \mu_{\varepsilon} \, \bar{u} \, \varepsilon + \tilde{p} \, \right] \,.
\end{eqnarray}
If we compare the expression above to (A10)-(12) in Broseta {\em et al.} \cite{Broseta87a},
we see that the prefactor of $\varepsilon^{\prime}(x)^2$ should read $\bar{u}^2$
and not $\bar{u}$.

The above free energy is minimized in two steps. First, the profile $\eta_0(x)$
is determined assuming that $\varepsilon(x) \!=\!0$. One finds that the bulk
value $\bar{\eta}$ ($\eta_{\ell} \!=\! \bar{\eta}$ and $\eta_{v} \!=\! -\bar{\eta}$)
is determined by ($\mu_{\eta} \!=\!0$ by symmetry):
\begin{equation}
\frac{1}{\omega} \ln \left( \frac{1 + \bar{\eta}}{1 - \bar{\eta}} \right) - \bar{\eta} = 0 \,,
\end{equation}
which has solutions for $\omega \!>\! \omega_c \!=\! 2$ ($\bar{\eta}_c \!=\! 0$).
The profile $\eta_0(x)$ is determined by solving:
\begin{equation}
\frac{\eta_0^{\prime}(x)^2}{4 \, (1 - \eta_0^2)} = \frac{(1+\eta_0)}{2 \omega} \, \ln(1+\eta_0)
+ \frac{(1 - \eta_0)}{2 \omega} \, \ln(1-\eta_0) - \frac{\eta_0^2}{4} + \tilde{p} \,.
\end{equation}
Second, using $\eta_0(x)$, the profile $\varepsilon_0(x)$ is determined.
In the bulk $\bar{\varepsilon} \!=\! 0$ which leads to the following expression
for $\mu_{\varepsilon}$:
\begin{equation}
\mu_{\varepsilon} = \frac{(1+\bar{\eta})}{2 \omega} \, \ln(1+\bar{\eta})
+ \frac{(1 - \bar{\eta})}{2 \omega} \, \ln(1-\bar{\eta}) 
- \frac{3 \nu + \chi}{4 \, (3 \nu - 1)} \, \bar{\eta}^2 \,.
\end{equation}
The profile $\varepsilon_0(x)$ is determined by solving:
\begin{eqnarray}
\frac{\bar{u}}{2} \, \varepsilon_0^{\prime\prime}(x) &=& \frac{(1+\eta_0)}{2 \omega} \, \ln(1+\eta_0)
+ \frac{(1 - \eta_0)}{2 \omega} \, \ln(1-\eta_0) \nonumber \\
&& + \frac{3 \nu \, K}{(3\nu-1)^2} \, \varepsilon_0 - \frac{(3 \nu + \chi)}{4 \, (3 \nu - 1)} \, \eta_0^2
- \mu_{\varepsilon} \,.
\end{eqnarray}
Again, comparing with (18) in Broseta {\em et al.} \cite{Broseta87a}, we see that
the additional factor of $\bar{u}$ in the expression for the free energy in Eq.(\ref{eq:F0_2})
leads to the presence of a factor $\bar{u}$ in front of $\varepsilon_0^{\prime\prime}(x)$.

With the help of these two profiles, the surface tension can then be calculated as:
\begin{equation}
\label{eq:A12}
\frac{\sigma \, \bar{\xi}^2}{k_{\rm B} T \, (\bar{u}/6)^{\frac{1}{2}}} = 1 - \Delta_1 - \bar{u} \Delta_2 \,.
\end{equation}
with 
\begin{eqnarray}
\Delta_1 &=& 1 - 2 \int\limits_{0}^{\bar{\eta}} \!\! d\eta
\left[ \frac{(1+\eta)}{2 \omega} \, \ln(1+\eta) + \frac{(1 - \eta)}{2 \omega} \, \ln(1-\eta)
- \frac{\eta^2}{4} \right]^{\frac{1}{2}} (1-\eta^2)^{-\frac{1}{2}} \,, \nonumber \\
\Delta_2 &=& \int\limits_{-\infty}^{\infty} \!\!\! dx
\left[ \frac{- \eta_0^{\prime}(x)^2}{8 \, (1 - \eta_0^2)}
+ \frac{(1 + \chi)}{8 \, (3 \nu - 1)} \, (\eta_0^2 - \bar{\eta}^2) \right] \varepsilon_0(x) \,.
\end{eqnarray}
One sees that $\Delta_1$ is straightforwardly calculated from the above integral,
but that for the calculations of $\Delta_2$ one needs to determine both density profiles
$\eta_0(x)$ and $\varepsilon(x)$. The expression for $\Delta_2$ also differs from
the one given in (29) in Broseta {\em et al.} \cite{Broseta87a}.


%
%

%
%
%


\begin{thebibliography}{99}

\bibitem{Bergfeldt}
Bergfeldt, K.; Piculell, L.; Linse, P. {\it J. Phys. Chem.} \textbf{1996}, 100, 3680.

\bibitem{Tolstoguzov}
Tolstoguzov, V. {\it Biotechnology Advances} \textbf{2006}, 24, 626.

\bibitem{Aarts}
Aarts, D.G.A.L.; van der Wiel, J.H.; Lekkerkerker, H.N.W. {\it J. Phys.: Condens. Matt.} \textbf{2003}, 15, 245.

\bibitem{Ryden}
Ryden, J.; Albertsson, P.-A. {\it J. of Coll. Interface Sci.} \textbf{1971}, 37, 219.

\bibitem{deHoog}
de Hoog, E.H.A.; Lekkerkerker, H.N.W. {\it J. Phys. Chem. B} \textbf{1999}, 103, 5274.

\bibitem{Scholten04}
Scholten, E.; Visser, J.E.; Sagis, L.M.C.; van der Linden, E. {\it Langmuir} \textbf{2004}, 20, 2292.

\bibitem{Scholten02}
Scholten, E.; Tuinier, R.; Tromp, R.H.; Lekkerkerker, H.N.W. {\it Langmuir} \textbf{2002}, 18, 2234.

\bibitem{Simeone}
Simeone, M.; Alfani, A.; Guido, S. {\it Food Hydrocolloids} \textbf{2004}, 18, 463.

\bibitem{vanPuyvelde}
van Puyvelde, P; Antonov, Y.A.; Moldenaers, P. {\it Food Hydrocolloids} \textbf{2002}, 16, 395.

\bibitem{Antonov}
Antonov, Y.A.; van Puyvelde, P.; Moldenaers, P. {\it International Journal of Biological
Macromolecules} \textbf{2004}, 34, 29.

\bibitem{Mitani}
Mitani, S.; Sakai, K. {\it Phys. Rev E.} \textbf{2002}, 66, 031604.

\bibitem{DeGennes}
de Gennes, P.G. {\it Scaling Concepts in Polymer Physics}; Cornell University Press: Ithaca, 1979.

\bibitem{Broseta86}
Broseta, D.; Leibler, L.; Lapp, A. {\it Europhys. Lett.} \textbf{1986}, 2, 733.

\bibitem{Broseta87a}
Broseta, D.; Leibler, L.; Kaddour, O.; Strazielle, C. {\it J. Chem. Phys.} \textbf{1987}, 87, 7248.

\bibitem{Broseta87b}
Broseta, D.; Leibler, L.; Joanny, J.-F. {\it Macromolecules} \textbf{1987}, 20, 1935.

\bibitem{Tromp} Tromp, R.H, Chapter 9 in {\it Structure and Functional Properties of Colloidal Systems},
Surfactant Science Series 146, Ed. Hidalgo-Alvarez, CRC Press: Boca Raton, 2010.

\bibitem{FH} Flory, P.F. {\it Principles of Polymer Chemistry}; Cornell University Press: Ithaca, 1953.

\bibitem{Lapp}
Lapp A.; Picot C.; Strazielle, C. {\it J. Phys. Lett.} \textbf{1985}, 46, L1031.

\bibitem{Tolman}
R.C. Tolman, J. Chem. Phys. \textbf{17}, 333 (1949).

\bibitem{Blokhuis92}
E.M. Blokhuis and D. Bedeaux, J. Chem. Phys. \textbf{97}, 3576 (1992).

\bibitem{Helfrich}
Helfrich, W. {\it Z. Naturforsch. C} \textbf{1973}, 28, 693.

\bibitem{Blokhuis93}
Blokhuis, E.M.; Bedeaux, D. {\it Mol. Phys.} \textbf{1993}, 80, 705.

\bibitem{Blokhuis06}
Blokhuis, E.M.; Kuipers, J. {\it J. Chem. Phys.} \textbf{2006}, 124, 074701.

\bibitem{Blokhuis13}
E.M. Blokhuis and A.E. van Giessen, {\tt arXiv:cond-mat/13046557}.

\end{thebibliography}
\end{document}